\documentstyle{article}
\input pz.sty
\input epsf.sty
\begin{document}

\PZhead{4}{26}{2006}{2 June}

\PZtitletl{The light curves of type I\lowercase{a} 
           Supernova 2004\lowercase{fu}}

\PZauth{D. Yu. Tsvetkov}
\PZinst{Sternberg Astronomical Institute, University Ave.13,
119992 Moscow, Russia; e-mail: tsvetkov@sai.msu.su}

\begin{abstract}
CCD $UBVRI$ photometry is presented for type Ia
supernova 2004fu in NGC 6949. The light and colour curves
are typical for this class of objects, absolute magnitude
at maximum and decline rate are in agreement with
the relationship between these parameters, established 
for SNe Ia.  
\end{abstract}

\begintext

SN 2004fu was discovered by Arbour (2004) on unfiltered CCD images taken 
with a 0.3-m Schmidt-Cassegrain reflector on November 4.87 (at mag
approximately 16.8) and 5.81 (mag approximately 16.5). The new
object was located at 
$\alpha  = 20\hr35\mm11\sec.54, \delta = +64\deg48\arcm25\arcs.7$
(equinox 2000.0), which is $29\arcs$ east and $15\arcs$ north of the 
center of the
Sc galaxy NGC 6949.  

Modjaz et al. (2004) report that a spectrum (range 350-740 nm)
of SN 2004fu, obtained on November 10.17
UT with the F. L. Whipple Observatory 1.5-m telescope, shows it to be a 
type-Ia supernova, several days
before maximum light. 

\medskip

We started photometric observations of SN 2004fu immediately 
after discovery, on November 6, with 60-cm reflector of 
Crimean Observatory of Sternberg Astronomical Institute (C60)
equipped with Roper Scientific VersArray1300B CCD camera,
and continued monitoring nearly each night until November 25. On 
November 22 images were
obtained at 38-cm reflector of Crimean Astrophysical Observatory (C38)
with Apogee AP-47p CCD camera. From 2004 December 29 until 2005 March 26
we observed SN 2004fu on four nights at 70-cm reflector in Moscow (M70)
with AP-47p CCD camera, and on 2005 May 4 last images were 
obtained at C60.  

\medskip

The reductions and photometric techniques were described by Tsvetkov (2006)
and Tsvetkov et al. (2006). The image of SN 2004fu with comparison
stars is shown in Fig.~1. Supernova is far away form the center
of the galaxy and from the spiral arms, and the galaxy background
does not affect the measurements.  
The magnitudes of the comparison stars are
presented in Table 1, and photometry of SN is reported in Table 2.

\medskip

The light curves are shown in Fig. 2, where we also plotted the
data obtained by amateur observers in the $BVR$ bands reported 
at SNWeb.\PZfm
\PZfoot{http://www.astrosurf.com/snweb2/2004/04fu/04fuMeas.htm}
Their data in $B$ are in good agreement with our results, while
in the $V$ and $R$ bands some systematic differences can be found. 
We achieved excellent photometric coverage on premaximum rise and first few 
days of brightness decline, but later only 
sporadic observations were obtained.
We can derive dates and magnitudes of maximum light: $B_{max}=15.87$ on
JD 2453326.3 (November 16.8), $V_{max}=15.42$ on JD 2453327.2, 
$R_{max}=15.18$         
on JD2453327.4. The date of maximum light in $I$ is
difficult to estimate with the same accuracy, as the 
light curve is quite flat near maximum, with $I_{max}=15.19$.
Using the data by amateurs,
we can derive the decline rate parameter $\Delta m_{15}(B)=1.3$.
We fitted the data with the light curves of the well studied SN Ia
1994D (Richmond et al., 1995, Altavilla et al., 2004) with 
nearly the same value of $\Delta m_{15}(B)$. The agreement between
the light curves of SN 2004fu and 1994D is nearly perfect in the $B$
and $V$ bands near maximum, while in $R$ and $I$ some differences are
evident. At late stages the difference between these objects is
prominent in all bands except $I$, but the late light curves are
not well sampled by our observations to allow definite conclusions. 

\medskip

The colour curves for SN 2004fu are shown in Fig. 3 and are compared
to those for SN 1994D. 
We see very good agreement between SN 2004fu and SN 1994D in $(B-V)$  
and some differences in $(V-R)$. We shifted the 
$(U-B), (B-V)$ and $(V-R)$
colour curves of SN 1994D by 0.5, 0.41 and 0.25 mag, respectively.
As the light of SN 1994D suffered negligible extinction both in the
Galaxy and in the host galaxy (Altavilla et al., 2004), these values
represent the total colour excess for SN 2004fu. Comparing these
data with the estimates of galactic extinction from Schlegel et al.
(1998) $A_U=2.10, A_B=1.67, A_V=1.28, A_R=1.03$ we conclude that 
all extinction towards SN 2004fu originated in the Galaxy. 
Taking these values for extinction and distance modulus 
$\mu$=33.19 from the LEDA database, 
we derive absolute magnitudes at maximum light:
$M_B=-18.99, M_V=-19.05, M_R=-19.04$. These values are 
slightly fainter than average luminosity for SNe Ia and are
in good agreement
with the relationship between decline rate parameter and absolute 
magnitude for 
SNe Ia as presented by Phillips (2005) and Altavilla et al. (2004).    

\medskip

We conclude that SN 2004fu is a typical object for its class
regarding photometric behavior,
with rate of decline after maximum slightly faster, and maximum
luminosity lower than average, in accordance with the relationship 
between these parameters for SNe Ia.      

\medskip

{\bf Acknowledgements:}
This research has made use of the Lyon-Meudon Extragalactic Database
(LEDA). The author is grateful to I.M.Volkov for help in the
observations. The work was partly supported by RFBR grant 05-02-17480.

\references

Altavilla, G., Fiorentino, G., Marconi, M., et al., 2004,
{\it MNRAS}, {\bf 349}, 1344

Arbour, R., 2004, {\it IAU Circ.}, No. 8428

Modjaz, M., Challis, P., Kirshner, R., Westover, M., 2004,
{\it IAU Circ.}, No. 8436

Richmond, M.W., Treffers, R.R., Filippenko, A.V., et al., 1995,
{\it Astron. J.}, {\bf 109}, 2121 

Phillips, M.M., 2005, {\it ASP Conf. Ser.,} {\bf 342}, 211,
in 1604-2004: Supernovae as cosmological lighthouses, M.Turatto et al. eds.

Schlegel, D., Finkbeiner, D., Davis, M., 1998, {\it Astrophys. J.},
{\bf 500}, 525 

Tsvetkov, D.Yu., 2006, {\it Permennye Zvezdy (Variable Stars)},
{\bf 23}, No. 3

Tsvetkov, D.Yu., Volnova, A.A., Shulga, A.P., et al., 2006,
{\it Astron. \& Astrophys. (submitted)}, astro-ph/0605184

\endreferences

\PZfig{12cm}{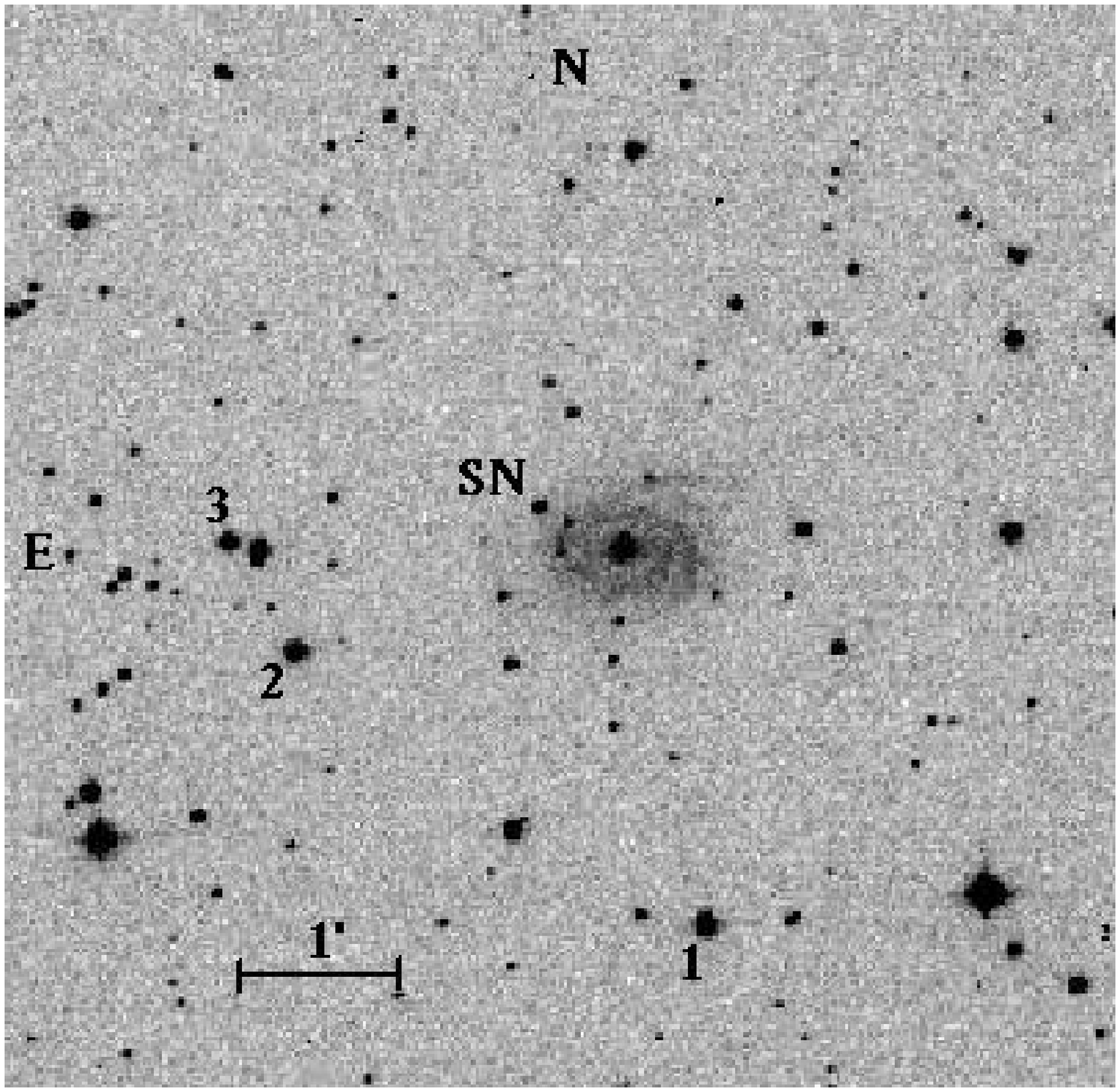}{SN 2004fu in NGC 6949 with comparison
stars}

\PZfig{12cm}{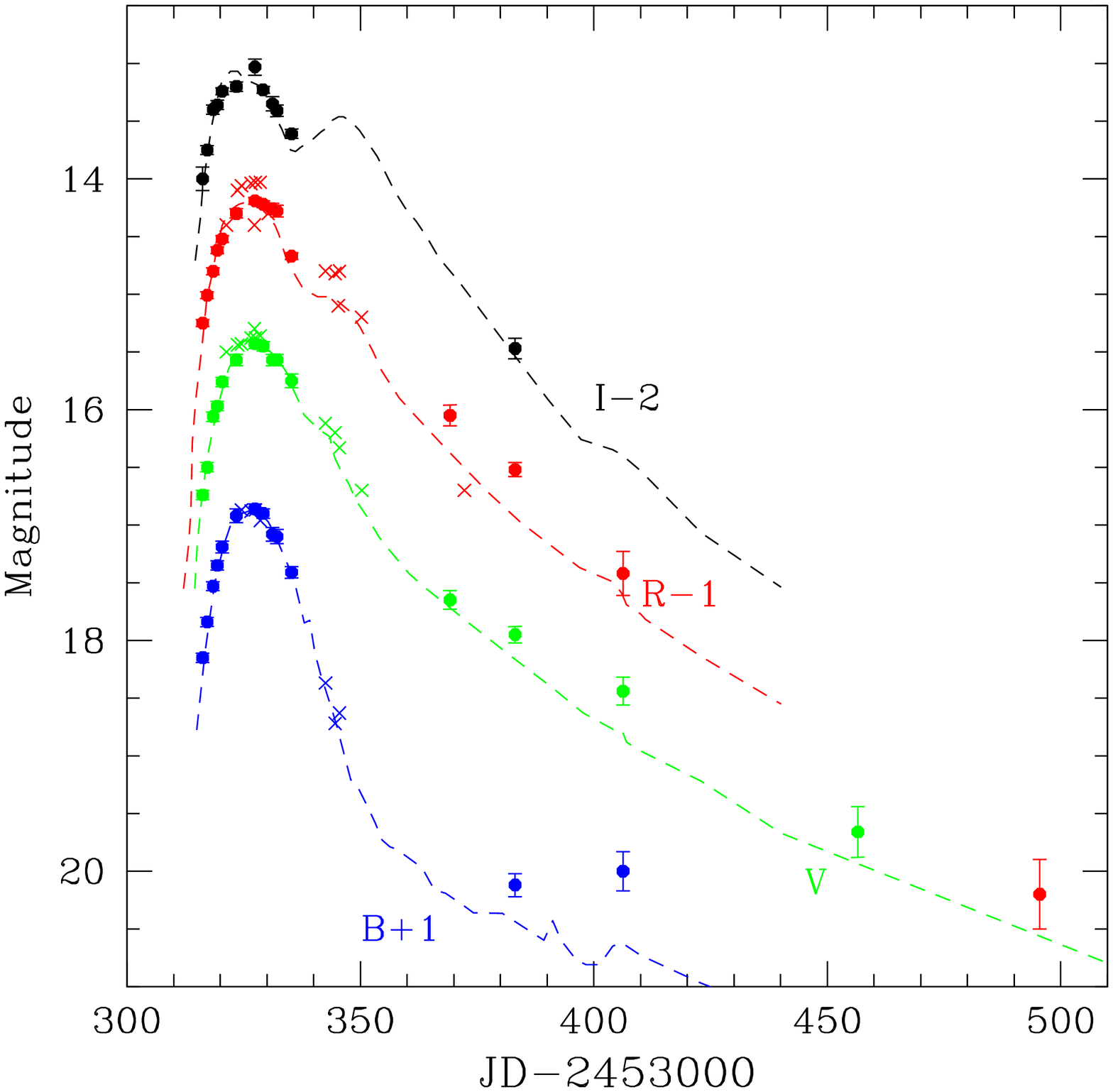}{$BVRI$ light curves of SN 2004fu,
showing our photometry (dots) and that of
amateur observers (crosses). The dashed lines are the light curves of 
SN 1994D}

\PZfig{12cm}{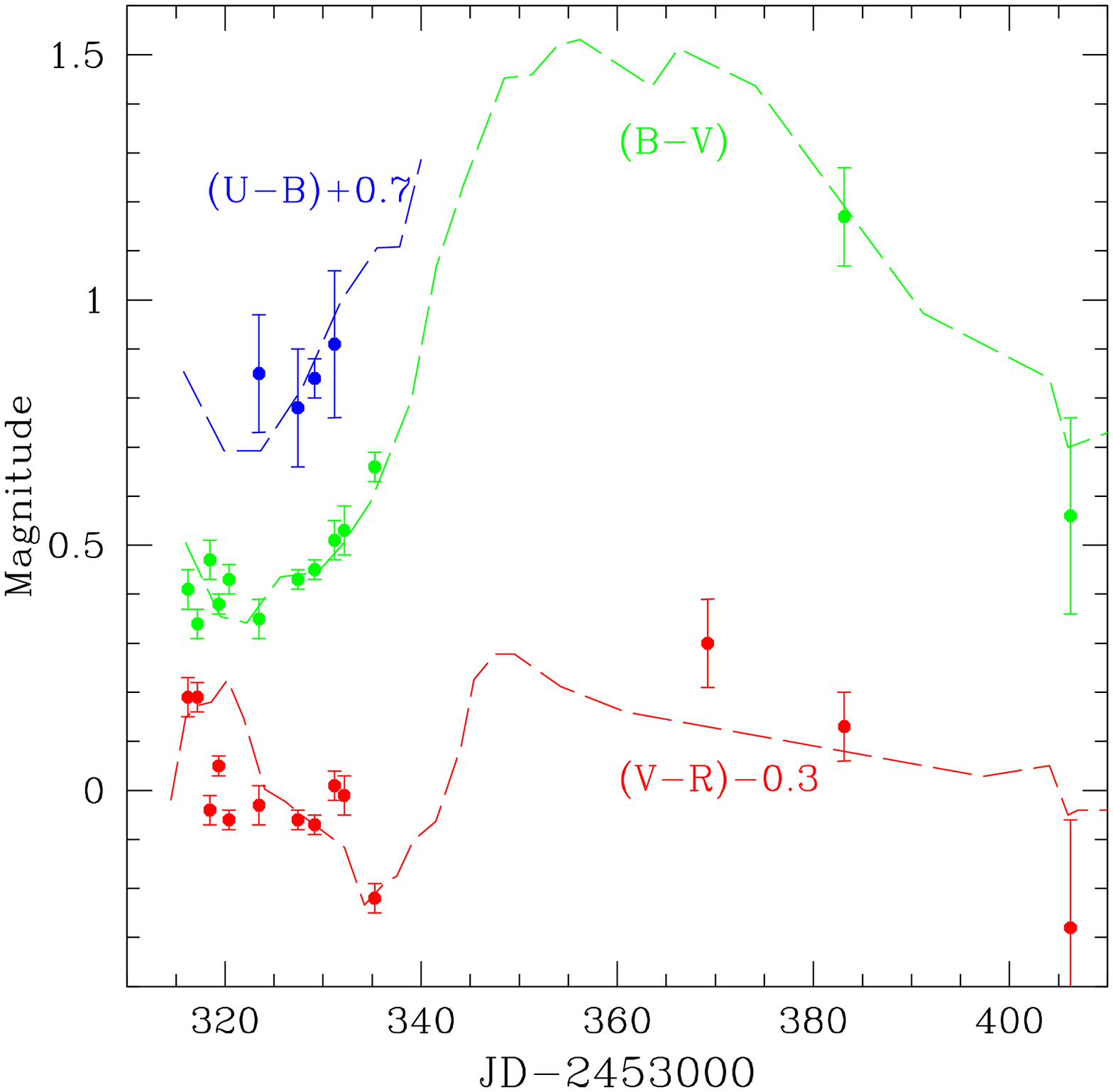}{$(U-B), (B-V)$ and $(V-R)$ colour curves 
of SN 2004fu.
The dashed lines are the colour curves of 
SN 1994D}

\newpage
\begin{table}
\caption{Magnitudes of comparison stars}\vskip2mm
\begin{tabular}{ccccccccccc}
\hline
Star & $U$ & $\sigma_U$ & $B$ & $\sigma_B$ & $V$ & $\sigma_V$ & $R$ & $\sigma_R$ 
& $I$ & $\sigma_I$
\\
\hline
1  & 14.66 & 0.06 & 14.59 & 0.03 & 13.85 & 0.04 & 13.34 & 0.02  & 12.92 & 0.02 \\
2  & 16.00 & 0.04 & 15.25 & 0.04 & 13.96 & 0.03 & 13.19 & 0.03  & 12.55 & 0.02 \\
3  & 15.73 & 0.07 & 15.25 & 0.04 & 14.22 & 0.04 & 13.59 & 0.03  & 13.12 & 0.02 \\
\hline
\end{tabular}
\end{table}

\begin{table}
\caption{Photometry of SN 2004fu}\vskip2mm
\begin{tabular}{cccccccccccc}
\hline
JD 2453000+ &$U$ & $\sigma_U$ &  $B$ & $\sigma_B$ & $V$ & $\sigma_V$ & $R$ & $\sigma_R$ & 
$I$ & $\sigma_I$ & Tel.\\
\hline
  316.22&       &     &  17.15& 0.04&  16.74& 0.04 &  16.25& 0.03&  16.00& 0.10& C60\\
  317.19&       &     &  16.84& 0.04&  16.50& 0.04 &  16.01& 0.03&  15.75& 0.04& C60\\
  318.45&       &     &  16.53& 0.04&  16.06& 0.04 &  15.80& 0.03&  15.40& 0.04& C60\\
  319.35&       &     &  16.35& 0.04&  15.97& 0.04 &  15.62& 0.03&  15.36& 0.04& C60\\
  320.41&       &     &  16.19& 0.05&  15.76& 0.04 &  15.52& 0.03&  15.24& 0.03& C60\\
  323.47&  16.07& 0.12&  15.92& 0.06&  15.57& 0.05 &  15.30& 0.04&  15.20& 0.04& C60\\ 
  327.42&  15.94& 0.12&  15.86& 0.04&  15.43& 0.04 &  15.19& 0.03&  15.03& 0.07& C60\\ 
  329.15&  16.04& 0.05&  15.90& 0.04&  15.45& 0.04 &  15.22& 0.03&  15.23& 0.03& C60\\ 
  331.17&  16.29& 0.15&  16.08& 0.06&  15.57& 0.05 &  15.26& 0.05&  15.35& 0.06& C60\\  
  332.15&       &     &  16.10& 0.06&  15.57& 0.05 &  15.28& 0.05&  15.41& 0.05& C38\\
  335.25&       &     &  16.41& 0.05&  15.75& 0.06 &  15.67& 0.03&  15.61& 0.04& C60\\
  369.22&       &     &       &     &  17.65& 0.08 &  17.05& 0.09&       &     & M70\\
  383.15&       &     &  19.12& 0.10&  17.95& 0.07 &  17.52& 0.06&  17.47& 0.09& M70\\
  406.21&       &     &  19.00& 0.17&  18.44& 0.12 &  18.42& 0.19&       &     & M70\\
  456.54&       &     &       &     &  19.66& 0.22 &       &     &       &     & M70\\
  495.49&       &     &       &     &       &      &  21.20& 0.30&       &     & C60\\
\hline
\end{tabular}
\end{table}
\end{document}